# COMPUTATIONAL STUDY OF ATMOSPHERIC TRANSFER RADIATION ON AN EQUATORIAL TROPICAL DESERT (LA TATACOA, COLOMBIA)

*Camilo Delgado-Correal, Jorge Hernández, Gabriel Castaño*
Remote Sensing Group-GEIPER (NIDE), Universidad Distrital FJC

## ABSTRACT

Radiative transfer models explain and predict interaction between solar radiation and the different elements present in the atmosphere, which are responsible for energy attenuation. In Colombia there have been neither measurements nor studies of atmospheric components such as gases and aerosols that can cause turbidity and pollution. Therefore satellite images cannot be corrected radiometrically in a proper way. When a suitable atmospheric correction is carried out, loss of information is avoided, which may be useful for discriminating image land cover. In this work a computational model was used to find radiative atmospheric attenuation (300∼1000nm wavelength region) on an equatorial tropical desert (La Tatacoa, Colombia) in order to conduct an adequate atmospheric correction.

## I. INTRODUCTION

Atmospheric correction is important for remote sensing because the interpreter can better discriminate images' elements. Atmospheric attenuation causes a considerable distortion of information received by sensors [1]. When satellite images are corrected, their quality is better and then it is possible to analyze and improve observation of land cover. This is useful in geology, study of natural resources, agricultural applications, and others. In order to understand and predict interaction between sunlight and different atmosphere constituents, the Simple Model of Atmospheric Radiative Transfer of Sunshine (SMARTS) was used. This model allows understanding more easily interaction between sunlight and atmosphere's components that cause scattering, absorption and transmission [2].

SMARTS is a single spectral model written in FORTRAN code predicts direct, diffuse and global surface radiation over small band increments ranging from 280 to 4000 nm including UV intervals, visible bands and near infrared. The SMARTS radiative transfer model was developed by Dr. Christian Gueymard (1995) and is distributed by the U.S. National Renewable Energy Laboratory (NREL) [3]. Its operation is based on a large number of absorption and scatering coefficients that are used according to type of atmosphere. It should be noted that the model must be implemented in clear sky areas.

## II. MATERIAL AND METHODS

### *A. **Study area.***

La Tatacoa is located at south of Colombia ($\boldsymbol{\varphi} = 3^{\circ}\,48'9"\,N\,;\ \boldsymbol{\lambda} = 75^{\circ}11'54''\,W$). It is an arid area with 330 $km^2$of extension. According to Caldas-Lang system classification this desert has erosive features with dry gorges. La Tatacoa is a warm dry desert belonging to a tropical dry forest. It is characterized as a plain area with some elevations in the northeast. Annual precipitation is around 1000 mm, average temperature is 28º C and exposure to sunshine is about 5.5 hours per day. There is reduced vegetation life, besides animals must adapt to high temperatures [4].

### *B. **Analysis of atmospheric information***

In order do to an atmospheric attenuation analysis, climate information from weather stations close to the study area was collected. Additional data taken from another images was used in order to execute the SMARTS model.

This study used weather monthly average data from Villavieja and San Alfonso stations run by the Instituto de Hidrologia, Meteorologia y Estudios Ambientales de Colombia-IDEAM. According to this information monthly sunshine varies between 130 to 170 hours. Maximum values occurred from June to September and lowest from February to April. Temperature greatest records where in July and August when relative humidity is low; it causes clear sky.

This desert rainfall pattern shows summer and winter periods repeated every three months where precipitation range varies from 20 to 170 m. Higher rainfall months are those with lower sunshine [4].

### C. ***Implementation of the radiative transfer model (SMARTS).***

SMARTS code is composed of 19 (nineteen) cards (Figure 1) that permit user to enter input data, such as name project, atmospheric pressure, type of atmosphere and principal features, pollution, water vapor and other atmospheric attenuation factor. Moreover, *Solar Geometry* (card 17) values must be introduced in order to achieve a better computational prediction of radiative attenuation. Solar radiation flux varies during the day.

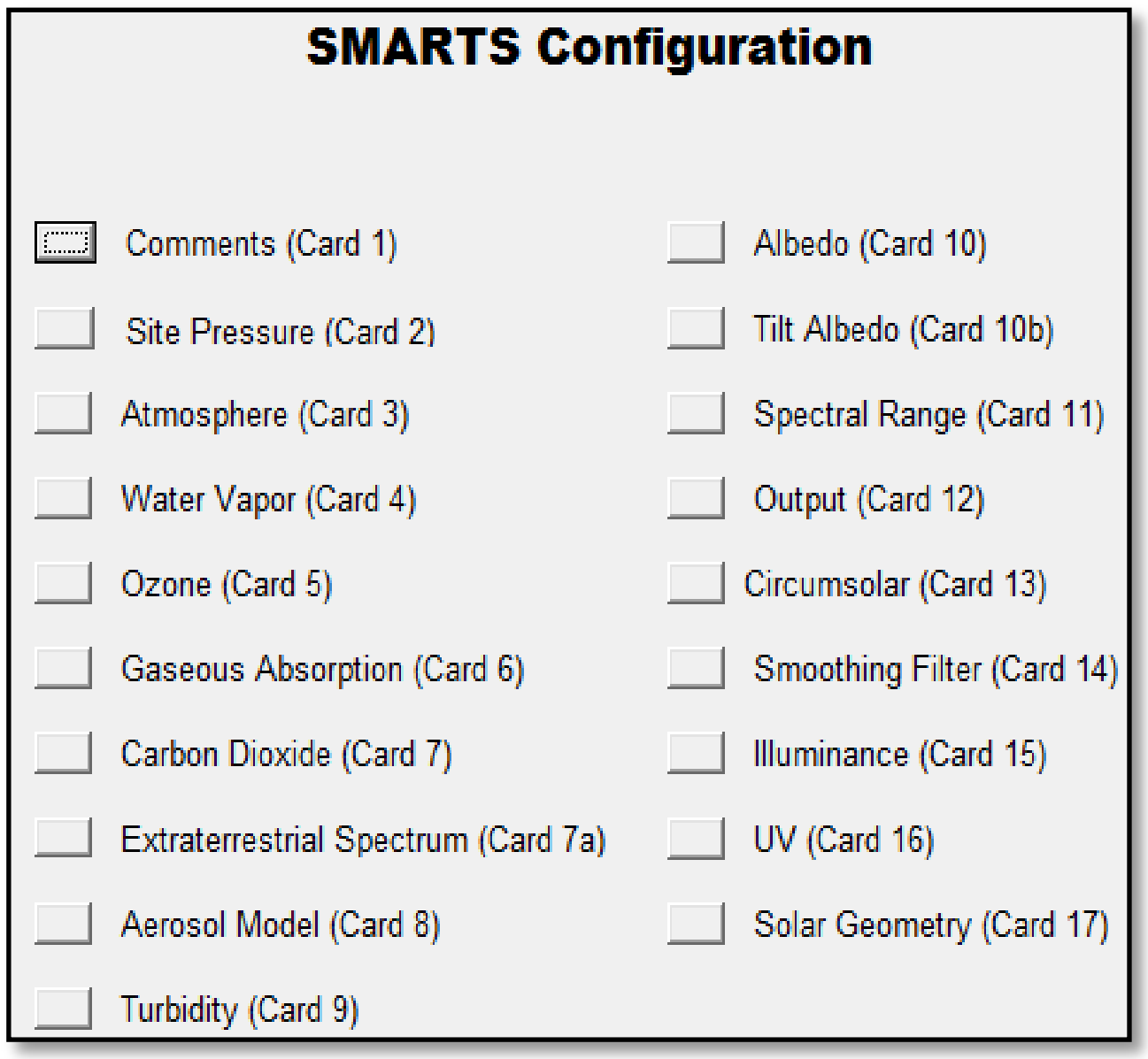

Figure 1. Input Data Setup.

The model also allows configuring output data which are always in text format, organized into columns according to the number of variables selected by user. TABLE 1 shows some of the parameters used in our current research:

TABLE 1.
Parameters used in smarts model.

| Parameter | Value |
|---|---|
| **Ground-level Air Temperature** | 31.3º C |
| **Ground –level Relative Humidity (%)** | 60 |
| **Average Daily Temperature** | 28.4 ºC |
| **Ozone vertical column** | 0.264 atm-cm |
| **Annual Precipitation** | ≈1100 mm |

Regional albedo (Figure 2), is one of the most important elements of SMARTS model. Dry clay soil albedo option was chosen in the regional albedo card because of the predominate conditions of the study area. This variable is essential to perform the subsequent analysis.

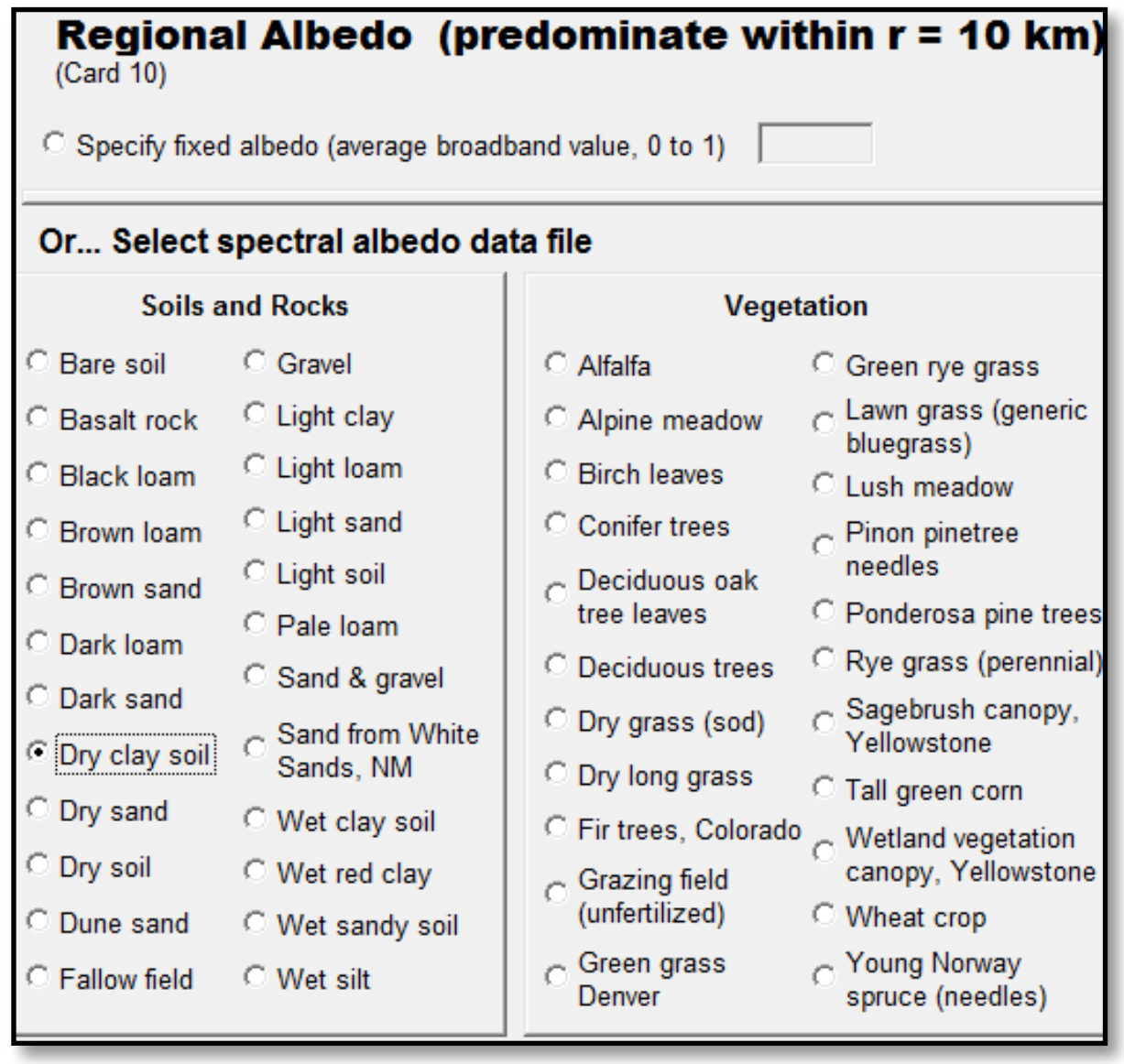

.

Figure 2. Some regional albedo in card (number 10).

### D. ***Physical principles using SMARTS model for atmospheric correction.***

The atmosphere is the essential element that disturbs solar radiation flux causing three kinds of alteration: absorption as a function of wavelength, scattering in certain spectral bands and emission caused by atmospheric temperature. Figure 3. illustrates interaction of sunlight and atmosphere. This allows users to understand the standard equation of surface reflectance $\boldsymbol{\rho_\lambda}$ (1) that is necessary to perform atmospheric correction of satellite images [5].

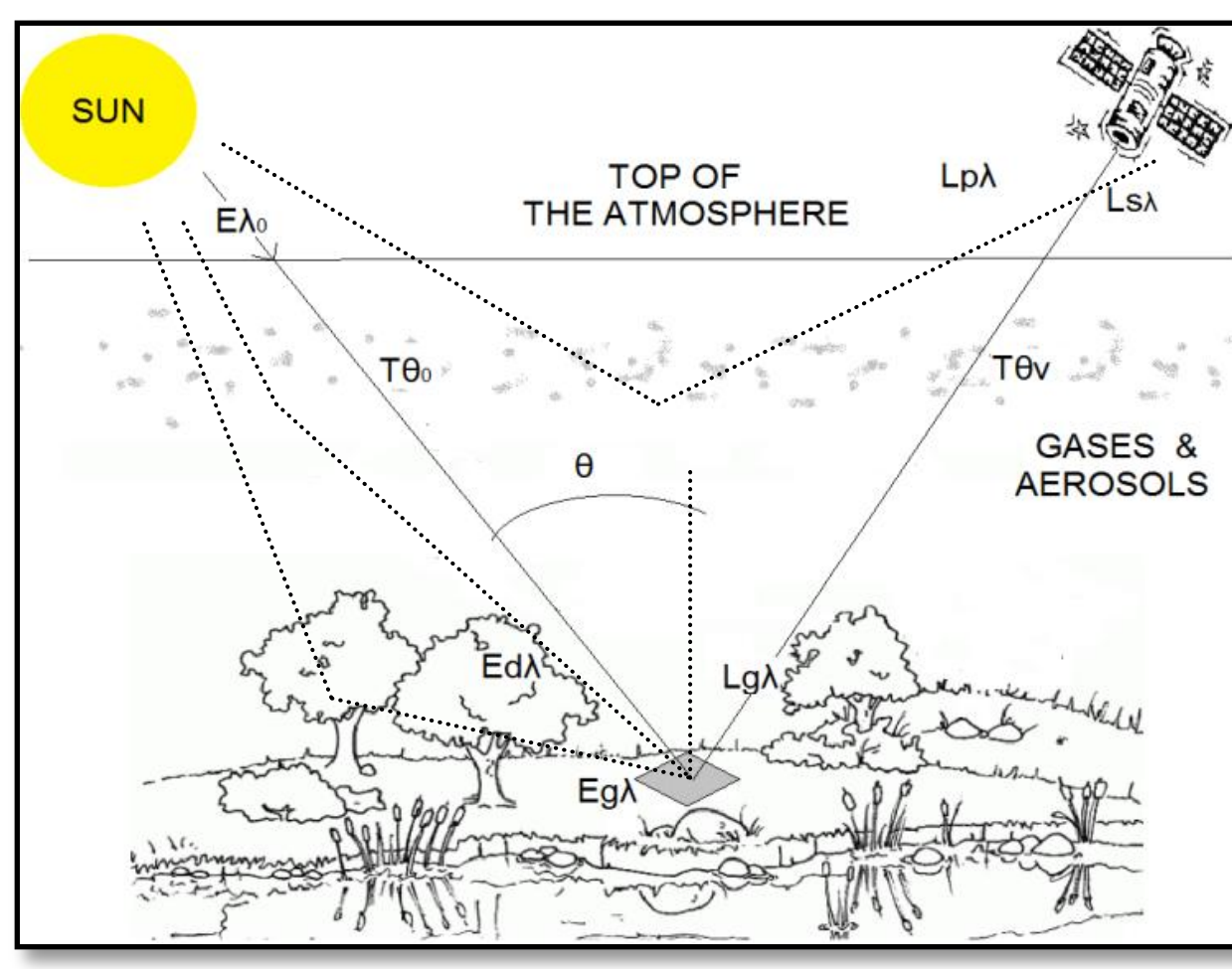

Figure 3. Interaction between solar radiation and atmosphere.

$$\boldsymbol{\rho_\lambda = \frac{\pi(L_{g\lambda})D}{E_{0\lambda}cos\theta_0 T_{\theta 0} + Ed_\lambda}} \qquad (1)$$

Where: $\boldsymbol{L_{g\lambda} = (L_{s\lambda} - L_{p\lambda}) / T_{\theta v}}$

At-surface spectral reflectance $\boldsymbol{\rho_\lambda}$ can be calculated as a function of: $L_{g\lambda}$= at-surface spectral radiance ($Wm^{-2}sr^{-1}\mu m^{-1}$); D=correction factor due to the change in the earth sun-distance (measure in astronomical units), $E_{0\lambda}$=spectral irradiance at the top-of-atmosphere (TOA) ($Wm^{-2}sr^{-1}\mu m^{-1}$), $T_{\theta 0}$= descending transmissivity, $\theta_0$= solar incident angle (or solar zenith angle) according to the ground surface slope, $Ed_\lambda$= diffuse spectral irradiance, $L_{s\lambda}$= at-satellite spectral radiance ($Wm^{-2}sr^{-1}\mu m^{-1}$), $L_{p\lambda}$= atmospheric path radiance and $T_{\theta v}$= ascending transmissivity.

The SMARTS model is useful to obtain and deduce parameters of in the surface reflectance equation such as diffuse spectral irradiance, descending transmissivity and TOA irradiance. In addition, SMARTS assumes values of the spectral surface radiance according to the predominant type of albedo within 100 m and 100 km. Furthermore, variables such as correction factor (D) and the sun inclination gle $\theta_0$ can be found in the image's metadata or can be determined with a solar calculator. At-surface spectral radiance and surface spectral reflectance are being contrasted with measurements acquired in situ using a spectroradiometer.

### *E. Our method.*

Nowadays, our team is doing an atmospheric correction of satellite images according to the following methodological framework proposed in Figure 4.

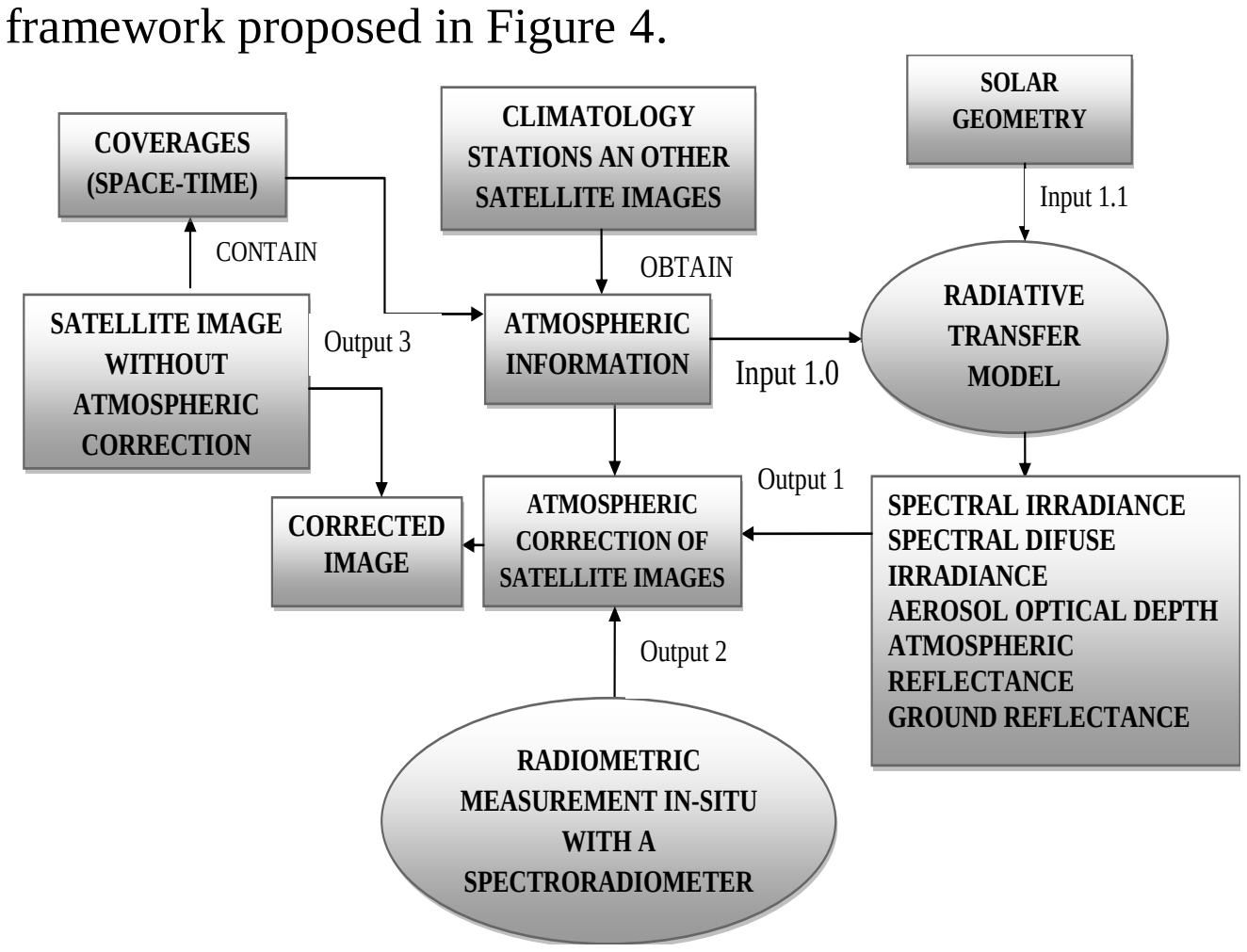

Figure 4. Methodological diagram for atmospheric correction to satellite images.

The basic procedure roughly consists of the following steps: First, a passive multispectral sensor captures the image from a specific area (raw information). Second, authors make a practical advance using climatology data or other satellite images. Third, the SMARTS model is executed using the information mentioned above in order to obtain the necessary variables to solve the surface reflectance equation. Fourth, those results are compared with radiometric measurements in-situ [6] (currently in course). Finally, the corrected image is obtained.

## III. RESULTS AND DISCUSSION

In order to evaluate the usefulness of the above proposed method to acquire corrected images, the analysis shown in Figure 5. (a), (b), (c) have been carried out.

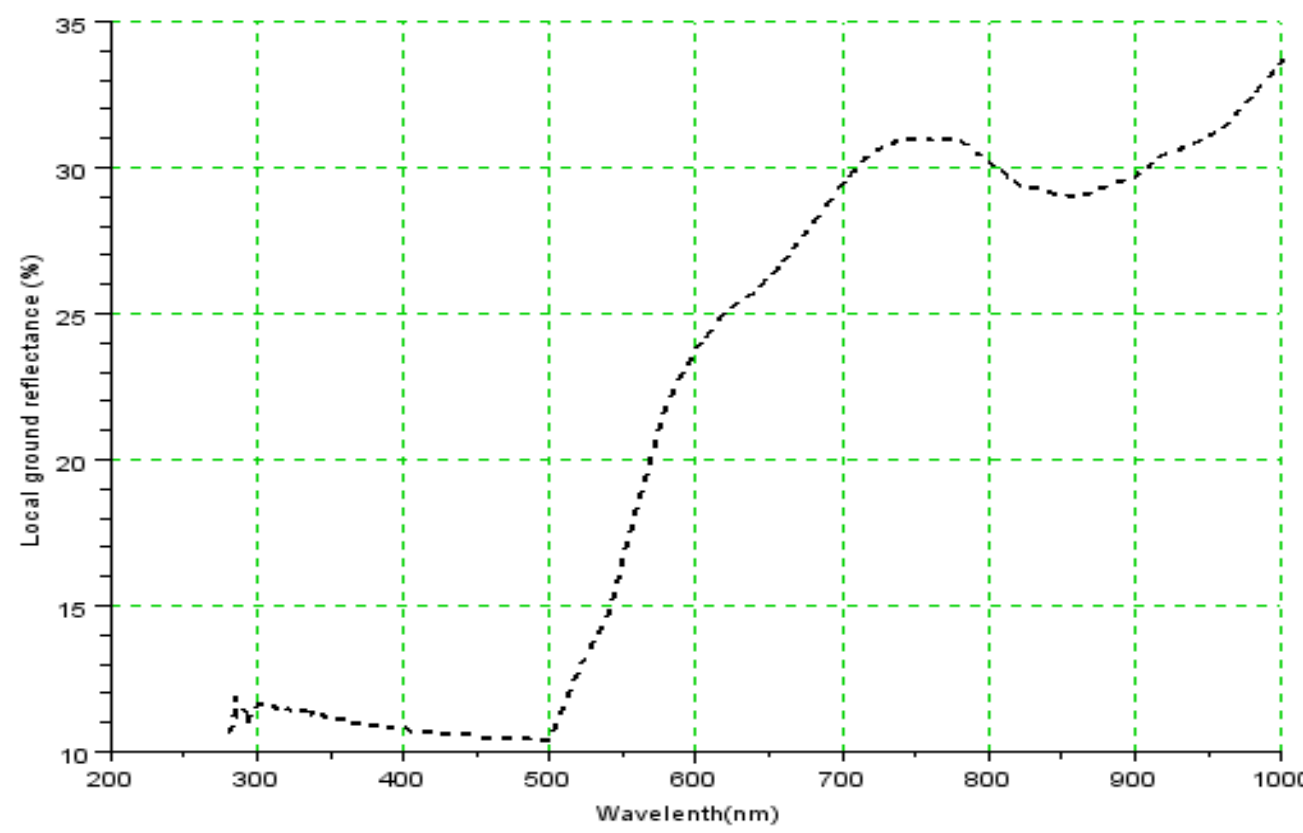

(a)

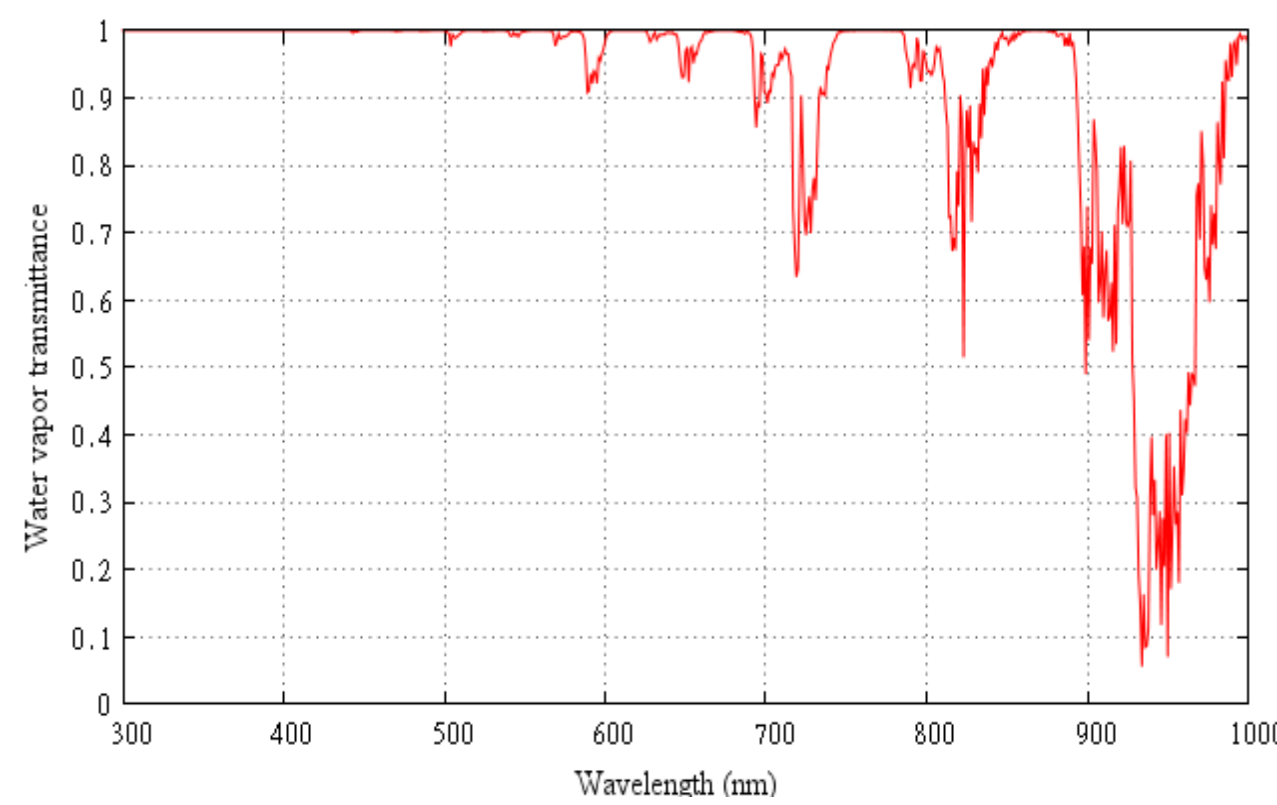

(b)

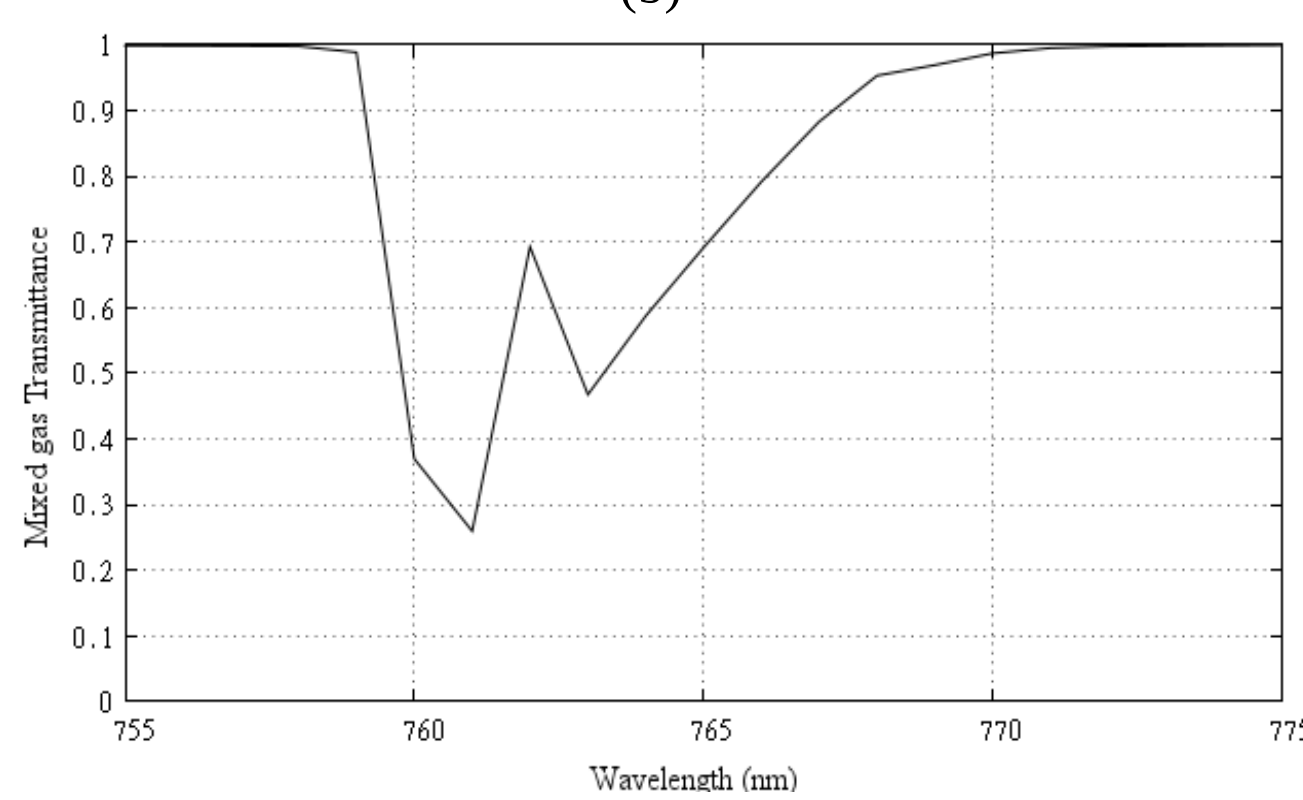

(c)

Figure 5. Computational prediction of radiative attenuation in La Tatacoa desert atmosphere. (a) Local ground reflectance; (b) Water vapor Transmittance; (c) Mixed gas-transmittance.

Factors that modify surface reflectance at La Tatacoa desert are: wetness content (closely linked to ground texture), soil roughness and iron oxide content. Its clay soil are usually poorly drained what causes low ground reflectance also decreases in visible ranges due to soil roughness and iron oxide content. Comparing Figure 5(a) with reflectance behavior of dry clay soil included in SMARTS code [3], authors found a similar absorption pattern in 800- 950nm spectral range. This result allows assuming La Tatacoa desert has dry clay soil.

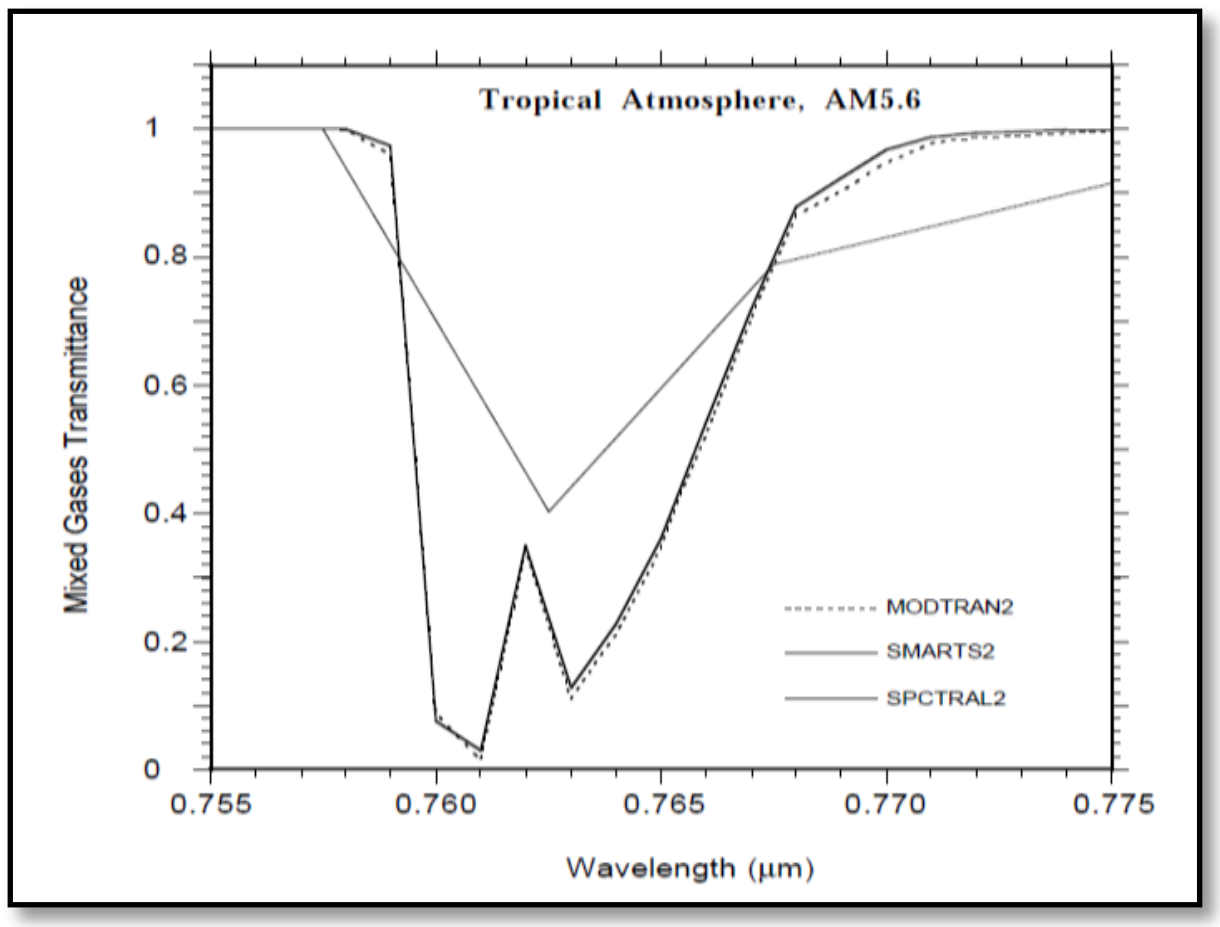

Figure 6. Mixed gas Transmittance in the visible and an air mass of 5.6 [2].

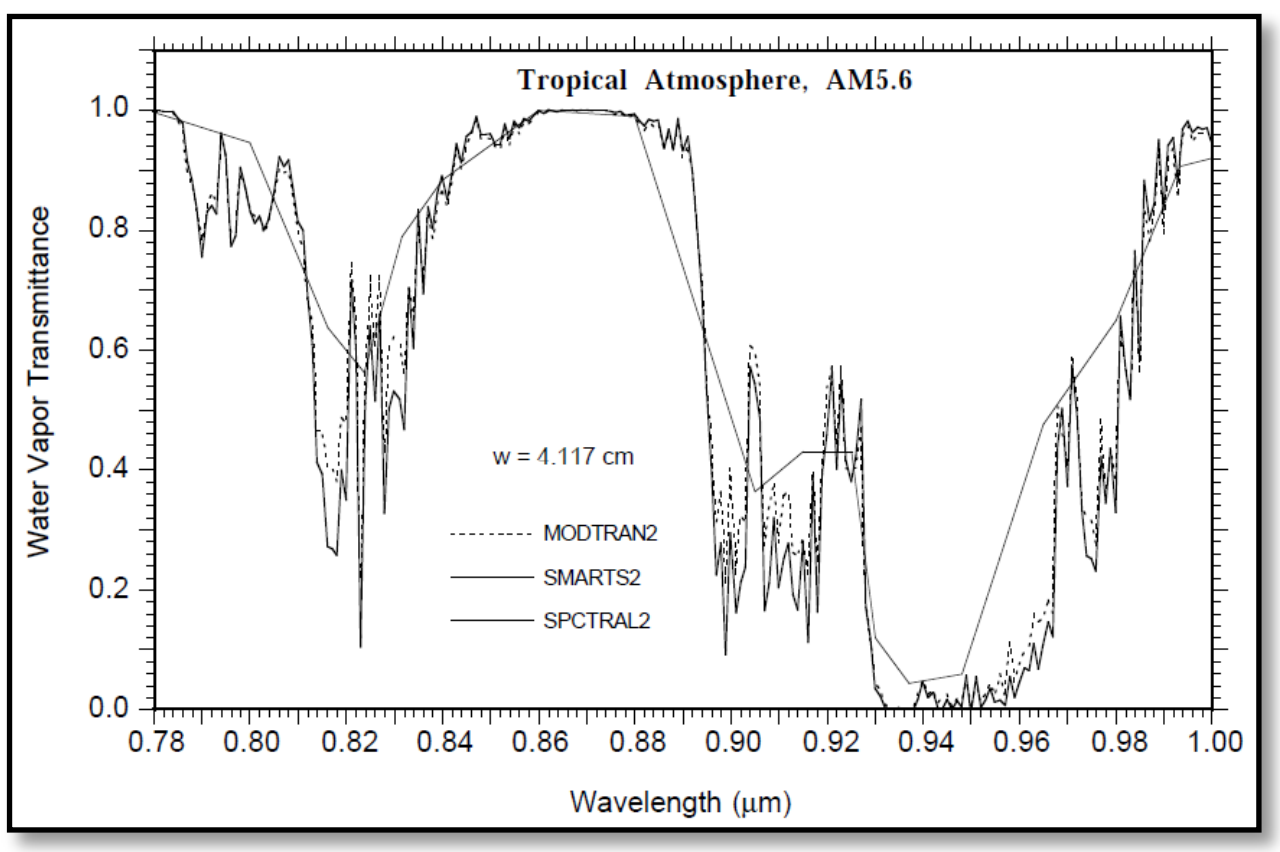

Figure 7. Water vapor transmittance for a Tropical reference atmosphere and an air mass of 5.6 [2].

The values of mixed gas transmittance obtained in the visible spectral range shows a considerable absorption in 760∼770nm wavelength region. It means solar radiation is being attenuated. Figure 6. shows values of mixed gases transmittance for a standard tropical Atmosphere proposed by C. Gueymard. According to our computational prediction using SMARTS model (Figure 5(c)), there is a noticeable difference in 760~765nm wavelength region between both results. This variation occurs because of low gases concentration in La Tatacoa desert atmosphere and authors expect to verify this assumption doing a field measurement.

Water vapor is the principal absorb constituent of the atmosphere in near infrared bands. It can be seen from Figure 5(b) and 7 that have the same pattern, despite having clear differences in 940∼950nm wavelength region. Water vapor transmittance values are higher in La Tatacoa desert than tropical reference atmosphere due to less annual precipitation, solar geometry, higher average daily temperature and elevated ground relative humidity at La Tatacoa desert.

## IV. CONCLUSION

La Tatacoa desert has particular land cover characteristics and atmospheric features that differentiate it from other desert areas. There is not enough information about surface reflectance of this desert, so it is necessary to use physically based methods to carry out atmospheric correction. Results obtained in this work using SMARTS will be used to do an adequate atmospheric correction of satellite imagery for our study area. We will do field measurements of ground reflectance at this desert to be used in standard atmospheric correction algorithms [7].

## ACKNOWLEDGMENT

We would like to acknowledge the technical support and discussion assistance of Ph.D. Ivan Lizarazo from Faculty of Engineering-Universidad Distrital Francisco Jose de Caldas.

## REFERENCES

[1] G. Rees, "Physical Principles of Remote Sensing", 2nd ed. Cambridge University Press, 2001.

[2] C.A. Gueymard, "Prediction and validation of cloudless shortwave solar spectra incident on horizontal, tilted, or tracking surfaces", Solar Energy, 82, pp. 260-271, 2008.

[3] C.A. Gueymard, "SMARTS, A Simple Model of the Atmospheric Radiative Transfer of Sunshine: Algorithms and Performance assessment", Professional Paper FSEC-PF-270-95. Florida Solar Energy Center, 1679 Clearlake Road, Cocoa, FL 32922, 1995.

[4] Universidad Surcolombiana, Faculty of Engeeniring, "Caracterización del área del desierto de la Tatacoa", Gobernación del Huila, Colombia, Dic 2006.

[5] D. Riano, E Chuvieco, J. Salas, y I. Aguado. "Assessment of different topographic corrections in Landsat-TM data for mapping vegetation types (2003)", Geoscience and Remote Sensing, IEEE Transactions on, vol, nº. 5, pp. 1056-1061, 2003.

[6] C. Delgado-Correal, J. E. Garcia, "Calibración Radiométrica In-Situ De Sensores Satelitales De Observación De La Tierra Utilizando Un Espectroradiómetro", Revista Colombiana de Física, Vol. 43, No.1, pp. 161-164, 2011.

[7] S. Liang, Hongliang Fang, M. Chen, "Atmospheric Correction of Landsat ETM+ Land Surface Imagery-Part I: Methods", IEEE Transactions on Geoscience and remote Sensing, 39, No.11, 2001.